\documentclass[conference]{IEEEtran}
\IEEEoverridecommandlockouts
\usepackage{cite}
\usepackage{amsmath,amssymb,amsfonts}
\usepackage{algorithmic}
\usepackage{bm}
\usepackage{graphicx}
\usepackage{textcomp}
\usepackage{xspace}
\usepackage{xcolor}
\usepackage{float}
\usepackage{cite}
\usepackage[final]{microtype}
\usepackage{mathtools}
\mathtoolsset{showonlyrefs=true}

\newcommand*{\QEDA}{\null\nobreak\hfill\ensuremath{\square}}%

\def\BibTeX{{\rm B\kern-.05em{\sc i\kern-.025em b}\kern-.08em
    T\kern-.1667em\lower.7ex\hbox{E}\kern-.125emX}}

\newcommand{\norm}[1]{\left\lvert\left\lvert#1\right\rvert\right\rvert}
\newcommand{\barpow}[1]{\left\lfloor#1\right\rceil}
\newcommand{\sign}[1]{\operatorname{sign}(#1)}

\newtheorem{thm}{Theorem}

\begin{document}

\title{A Discrete-Time Matching Filtering Differentiator*\\
\thanks{The authors sincerely thank CONACyT for the scholarship provided during this investigation to the student with No. CVU 555845, and to CINVESTAV for the provided resources.} 
}

\author{
\IEEEauthorblockN{
J.~E. Carvajal-Rubio}
\IEEEauthorblockA{\textit{Dept. of Electrical Engineering} \\
\textit{CINVESTAV-IPN, Guadalajara}\\
Zapopan, M\'exico \\
Jose.Carvajal@cinvestav.mx}
\and
\IEEEauthorblockN{J.~D. S\'anchez-Torres}
\IEEEauthorblockA{\textit{Dept. of Mathematics and Physics} \\
\textit{ITESO}\\
Tlaquepaque, M\'exico \\
dsanchez@iteso.mx}
\and
\IEEEauthorblockN{M. Defoort}
\IEEEauthorblockA{\textit{LAMIH, CNRS UMR 8201} \\
\textit{Polytechnic University of Hauts-de-France}\\
Valenciennes, France \\
michael.defoort@uphf.fr}
\\
\and
\IEEEauthorblockN{A.~G. Loukianov}
\IEEEauthorblockA{\textit{Dept. of Electrical Engineering}\\
\textit{CINVESTAV-IPN, Guadalajara}\\
Zapopan, M\'exico \\
louk@gdl.cinvestav.mx}
\and
\IEEEauthorblockN{M. Djemai}
\IEEEauthorblockA{\textit{LAMIH, CNRS UMR 8201} \\
\textit{Polytechnic University of Hauts-de-France}\\
Valenciennes, France \\
Mohamed.Djemai@uphf.fr}
}

\pretolerance=8000
\tolerance=8000

\maketitle

\begin{abstract}
This paper presents a time discretization of the robust exact filtering differentiator, a sliding mode differentiator coupled to filter, which provides a suitable approximation to the derivatives of some noisy signals. This proposal takes advantage of the homogeneity of the differentiator, allowing the use of similar techniques to those of the linear systems. As in the original case, the convergence robust exact filtering differentiator depends on the bound of a higher-order derivative; nevertheless, this new realization can be implemented with or without the knowledge of such constant.  It is demonstrated that the system's trajectories converge to a neighborhood of the origin with a free-noise input. Finally, comparisons between the behavior of the differentiator with different design parameters are presented.  

\end{abstract}


\section{Introduction}

Usually, a control law or an observer is designed in continuous-time, but it is implemented in a digital system. They are implemented under the assumption that the sampling time is small enough to preserve its continuous-time property. However, its properties can be lost or modified. Different methodologies have been proposed to obtain adequate realizations, aimed to preserve those properties of the continuous-time systems. Some examples are  Euler method, Exact discretization \cite{kazantzis1999time}, and implicit discretization \cite{brogliato2020_tutorial} to name a few.  

On the other hand, a differentiator allows to implement many applications, such as control laws based on derivatives of a signal, and estimation of unmeasured states and parameters \cite{kaveh2008blood,shtessel2007smooth,iqbal2010robust}. In \cite{Levant_HSMD} a homogeneous differentiator was proposed, it can estimate the first $n$ derivatives of a signal with a bounded $(n+1)$-th derivative. Moreover, it presented robustness to delays and bounded noises. Different time discretization methodologies has been used with the objective of preserve its accuracy and robustness \cite{Miki2014,koch2019discrete,barbot2020discrete,carvajal2019discretization,Stefan_dif}. Recently, a new robust exact filtering differentiator was presented in \cite{levant2019robust}, which improves the accuracy and presents the desirable properties of the differentiator presented in \cite{Levant_HSMD}. 

This paper's contribution is a new discrete-time differentiator and a demonstration of its convergence to a neighborhood of the origin. This discrete-time differentiator is based on the methodology presented in \cite{Stefan_dif} and the robust exact filtering differentiator \cite{levant2019robust}. This paper is organized as follows. In Section \ref{sec:Problem}, a summary of the differentiation problem is presented. In Section \ref{sec:differentiation}, the standard differentiator \cite{levant2019robust} and robust exact filtering differentiator \cite{levant2019robust} are introduced and compared. In Section \ref{sec:discretization}, the proposed discrete-time realization is given and analyzed. In Section \ref{sec:Results}, with the purpose of showing the performance of the new discrete-time differentiator, two simulations are presented with different selected parameters and conditions. In Section \ref{sec:conclusion}, the main results of the paper are summarized, and future work is presented.

\section{Problem statement and preliminaries.}\label{sec:Problem}

\subsection{Notation and Properties.}

Let $x\in \mathbb{R}$. The absolute value of $x$, denoted by $|x|$, is defined as $\left|x\right|=x$ if $x\geq 0$, and $\left|x\right|=-x$ if $x<0$. The set-valued function $\sign{x}$ is defined as $\sign{x}=\{1\}$ for $x>0$, $\sign{x}=\{-1\}$ for $x<0$, and $\sign{x}=\left[-1,1\right]$ for $x=0$. For $\gamma\geq0$, the signed power $\gamma$ of $x$ is defined as $\barpow{x}^{\gamma}=\left| x\right|^{\gamma}\sign{x}$, particularly, $\barpow{x}^0=\sign{x}$.

For any matrices $\bm{C}$, $\bm{D}\in \mathbb{R}^{n \times m}$ and any symmetric positive definite matrix $\bm{\Lambda} \in \mathbb{R}^{n\times n}$ the following inequality hold:

\begin{align}\label{eq:matrixprop}
    \begin{split}
        \bm{C}^T \bm{D} +  \bm{D}^T \bm{C} \leq \bm{C}^T \bm{\Lambda}\bm{C}+\bm{D}^T \bm{\Lambda}^{-1}\bm{D},
    \end{split}
\end{align}
this property can be found in \cite{Poznyak2008}.

\subsection{Problem statement.}

The objective of a differentiator is to obtain online the first $n$ derivatives of a function even if there is noise in the measurement. $f_0\left( t \right)$ represents this function, $f_0:\mathbb{R}\rightarrow \mathbb{R}$. $f_0(t)$ is assumed a function at least $(n+1)-th$ differentiable and its $n+1$ derivative is bounded by a known real number $L>0$, i.e., $|f_0^{\left( n+1 \right)}\left( t\right)|\leq L$. The input of the differentiator is defined as $f(t)=f_0(t)+\Delta\left(t\right)$ and $\Delta\left(t\right)$ correspond to the noise in the input. Additionally, it is also assumed that $\Delta\left( t \right)$ is a Lebesgue-measurable bounded noise with $|\Delta(t)|\leq \delta$ for a real number $\delta>0$, which can be unknown. 

To design a differentiator, a space state representation is used, it allows to compute the derivatives $f_0^{(1)}(t)$, $f_0^{(2)}(t)$, $\cdots$, $f_0^{(n)}(t)$. The state variables are defined as $x_i(t)=f_0^{(i)}(t)$ and $\bm{x}=\left[ \begin{array}{ccccc}
    x_0 & x_1 & x_2 & \cdots  & x_n
\end{array} \right]^T \in \mathbb{R}^{n+1}$. Therefore, one can obtain the following representation for the differentiation problem in the state space:
\begin{equation}\label{eq:difsys}
\begin{array}{lll}
\dot{\bm{x}}(t)&=&\bm{A}\bm{x}(t)+\bm{e}_{n+1}f_0^{(n+1)}(t)\\
f(t)&=&\bm{e}_1^T\bm{x}(t)+\Delta(t)
\end{array}
\end{equation}
with the canonical vectors $\bm{e}_{1}=\left[ \begin{array}{ccccc}
    1 & 0 & \cdots & 0 & 0 
\end{array} \right]^T$, $\bm{e}_{n+1}=\left[ \begin{array}{cccccc}
    0 & 0 & \cdots & 0 & 1 
\end{array} \right]^T$ and $\bm{A}=[\bm{0}_{(n+1)\times 1}\;\bm{e}_1\;\bm{e}_2\;\cdots\;\bm{e}_n]$, which is a nilpotent matrix of appropriate dimensions. Notice that the successive time derivatives of $f_0\left( t \right)$ can be obtained through the design of a state observer. 

\section{Differentiation} \label{sec:differentiation}

\subsection{Standard Differentiator}
With the purpose of obtain the first $n$ derivatives of $f_0\left(t\right)$, a continuous-time differentiator was proposed in \cite{Levant_HSMD} as: 

\begin{equation}\label{eq:diflev1}
\dot{\bm{z}}=\bm{A}\bm{z}+\bm{B}\bm{u}\left(\sigma_0-\Delta(t)\right)   
\end{equation}
where $\bm{u}\left(\sigma_0\right)=\left[\upsilon_{0,n}\left( \sigma_0 \right)\; \upsilon_{1,n}\left( \sigma_0 \right)\; \cdots \; \upsilon_{n,n}\left( \sigma_0 \right) \right]^T$, $\upsilon_{j,n}\left( \cdot \right)=-\lambda_{n-j}L^{\frac{j+1}{n+1}} \barpow{\cdot}^{\frac{n-j}{n+1}}$, $\bm{B}$ is the identity matrix of appropriate dimensions, $\sigma_j=z_j-x_j$ and $\bm{z}=\left[ \begin{array}{ccccc}
    z_0 & z_1 & z_2 & \ldots  & z_n
\end{array} \right]^T$ is the finite-time estimate of the state vector $\bm{x}$ using adequate  $\lambda_j>0$ (see \cite{levant2019robust }). Sequences of parameters $\lambda_j$ are presented in \cite{levant2019robust } for $n\leq 7$, but , they are not unique due to the fact that the sequences can be built for any $\lambda_0>1$ \cite{Levant_HSMD}. For instance, in \cite{Reichhartinger_lambda}, $\lambda_j$ is defined for $1\leq n\leq 10$. Since the function $ \barpow{z_0-f\left(t\right)}^{0}$ is discontinuous at $z_0=f$, the solutions of system \eqref{eq:diflev1} are understood in the Filippov sense \cite{filippov2013differential}. Under the above assumption with respect to $f_0(t)$, $\Delta(t)$, $L$ and $\lambda_j$, the standard differentiator \eqref{eq:diflev1} ensures the following precision
\begin{align}\label{eq:acc_standard_dif}
    \begin{split}
        |z_j-&f_0^{(j)}\left(t\right)|\leq \mu_j L^{\frac{j}{n+1}} \delta^{\frac{n+1-j}{n+1}} ,\;\mu_j>0,\\
        &j=0, 1, \cdots, n,
    \end{split}
\end{align}
which correspond to an asymptotically optimal accuracy \cite{levant2017}.

\subsection{Robust Exact Filtering Differentiator}

Although, differentiator \eqref{eq:diflev1} offers good performance when there exists a Lebesgue-measurable bounded noise $\Delta(t)$ such that $|\Delta(t)|\leq \delta$ with small average $\delta$, its performance becomes significantly reduced when $\delta$ is large. On the other hand, a bounded noise is a signal of filtering order $0$ and integral magnitude $\epsilon_0\geq 0$. Now, it is assumed that $\Delta(t)$ is presented as $\Delta(t)=\Delta_0(t)+\Delta_1(t)+\cdots+\Delta_{n_f}(t)$, where $\Delta_j(t)$ is a signal of the global filtering order $j$ and integral magnitude $\epsilon_j\geq 0$ with $j=0,1,\cdots, n_f$. More details can be founded in \cite{levant2019robust}. Note that a bounded noise signal satisfies the above assumption. In \cite{levant2019robust}, a new finite-time robust exact filtering differentiator has been proposed for those noises, with the following structure:
\begin{align}\label{eq:cont_dist_filt}
    \begin{split}
        \dot{w}_{j_f}&=-\lambda_{m+1-j_f} L^{\frac{j_f}{m+1}} \barpow{w_1}^{\frac{m+1-j_f}{m+1}}+w_{j_f+1} \\
         \dot{w}_{n_f}&=-\lambda_{n+1} L^{\frac{n_f}{m+1}} \barpow{w_1}^{\frac{n+1}{m+1}}+z_{0}-f\left(t\right) \\
        \dot{z}_{j_d}&=-\lambda_{n-j_d} L^{\frac{n_f+1+j_d}{m+1}} \barpow{w_1}^{\frac{n-j_d}{m+1}}+z_{j_d+1}\\
        \dot{z}_{n}&=-\lambda_{0} L \barpow{w_1}^{0}\\
        j_f&=1, 2, \cdots, n_f-1.  \;\;\;\;\;  j_d=0, 1, 2, \cdots, n-1.
    \end{split}
\end{align}
where $m=n+n_f$, $n_f\geq 0$, $n_f$ is the filtering order and the parameters $\lambda_j$ are selected as in \eqref{eq:diflev1}. $n_f$ of the differentiator \eqref{eq:cont_dist_filt} can be selected greater than the highest filter order of $\Delta_j(t)$. Furthermore, it is shown that differentiator \eqref{eq:cont_dist_filt} offers the following accuracy:
\begin{align}\label{eq:acc_robust_dif}
    \begin{split}
        |z_j-&f_0^{(j)}\left(t\right)|\leq \mu_j L\rho^{n+1-j},\;\;\mu_j>0,\;\;j=0, 1, 2, \cdots, n.\\
        \rho=&\max \left [ \left(\frac{\epsilon_0}{L} \right)^{\frac{1}{n+1}}, \left(\frac{\epsilon_1}{L} \right)^{\frac{1}{n+2}}, \cdots,  \left(\frac{\epsilon_{n_f}}{L} \right)^{\frac{1}{m+1}}\right].
    \end{split}
\end{align}

For a bounded noise, the accuracy \eqref{eq:acc_robust_dif} presents the structure of the accuracy \eqref{eq:acc_standard_dif}. The advantage of use the robust exact filtering differentiator \eqref{eq:cont_dist_filt} instead of standard one \eqref{eq:diflev1}, is that \eqref{eq:acc_robust_dif} is a better accuracy than \eqref{eq:diflev1}. Moreover, filtering differentiator \eqref{eq:acc_robust_dif} rejects unbounded noises with a small local average \cite{levant2019robust}. As in \cite{Stefan_dif}, for a free-noise case ($\Delta(t)=0$), the error system can be presented as:

\begin{align}\label{eq:cont_dist_filt_mat}
    \begin{split}
        &\left[ \begin{array}{cc}
         \dot{\bm{w}}  \\
         \dot{\bm{\sigma}} 
         \end{array} \right]=  \bm{E} \left[ \begin{array}{cc}
         \bm{w}  \\
         \bm{\sigma} 
         \end{array} \right]-\bm{e}_{m+1} f_0^{(n+1)}(t),\\
        &\bm{E}=\left[ \begin{array}{cccccc}
         -\lambda_{m} L^{\frac{1}{m+1}} |w_1|^{\frac{-1}{m+1}} & 1 & 0 &   \cdots & 0 \\
         -\lambda_{m-1} L^{\frac{2}{m+1}} |w_1|^{\frac{-2}{m+1}} & 0 & 1 &  \cdots & 0 \\
         \vdots &  \vdots & \vdots & \cdots & \vdots\\
         -\lambda_{1} L^{\frac{m}{m+1}} |w_1|^{\frac{-m}{m+1}} & 0 & 0 &  \cdots & 1 \\
         -\lambda_{0} L |w_1|^{-1} & 0 & 0 &  \cdots & 0  
         \end{array} \right],\\
         &\bm{e}_{m+1}=\left[ 0  \;\; \cdots \;\; 0 \;\; 1 \right]^T,
    \end{split}
\end{align}
where the dimensions of the matrix $\bm{E}$ and $\bm{e}_{n_f}$ are $(m+1)\times (m+1)$ and $(m+1) \times 1$ respectively, $\bm{w}=\left[w_1 \;\; w_2 \;\; \cdots \;\; w_{n_f}  \right]^T$ and $\bm{\sigma}=\left[\sigma_0 \;\; \sigma_1 \;\; \cdots \;\; \sigma_{n}  \right]^T$. The characteristic equation of $\bm{E}$ is $P(s)=s^{m+1}+\lambda_{m}L^{\frac{1}{m+1}}|w_1|^{\frac{-1}{m+1}} s^m+\lambda_{m-1}L^{\frac{2}{m+1}}|w_1|^{\frac{-2}{m+1}}s^{m-1}+\cdots+\lambda_0 L|w_1|^{-1}$, its roots can  be calculated by using the equation:

\begin{align}\label{eq:pol_conti}
    \begin{split}
        \left( |w_1|^{\frac{1}{m+1}}s \right)^{m+1}+\lambda_{m}L^{\frac{1}{m+1}}\left(|w_1|^{\frac{1}{m+1}} s\right)^m+\cdots+\lambda_0 L=0.
    \end{split}
\end{align}
Therefore, the $m+1$ roots $c_j$ of the characteristic equation of $\bm{E}$ can be calculated of the following polynomial:

\begin{align}\label{eq:pol_conti_b}
    \begin{split}
        Q(b)=b^{m+1}+\lambda_{m}L^{\frac{1}{m+1}} b^m+\cdots+\lambda_0 L.
    \end{split}
\end{align}
Then $c_j$ is calculated as $c_j=|w_1|^{\frac{-1}{m+1}} b_j$, where $b_j$ correspond to the roots of the polynomial \eqref{eq:pol_conti_b}. This result will be used in the Section \ref{timeDiscretization}.

\section{Discretization of the continuous-time systems}\label{sec:discretization}

Let us denote  the measurement time as $t_k$ and $x_{j,k}=x_j\left( t_k \right)$, $\bm{x}_k=\left[ x_{0,k}, \ldots, x_{n,k}\right]^T$. Then,

\begin{align}\label{eq:disc_lev1}
    \begin{split}
        x_{j,k+1}&=\sum_{l=j}^{n} \frac{\tau^{l-j}}{(l-j)!} x_{l,k}+ h_{j,k} (\tau)\\
        j&=0,1,2,\cdots, n.
    \end{split}
\end{align}
is a discrete-time representation of continuous-time system \eqref{eq:difsys}, where the sampling time is defined as $\tau=t_{k+1}-t_k$. It is obtained using Taylor series expansion with Lagrange's remainders \cite{firey1960remainder,Apostol1967}. If $f_0^{(n+1)}(t)$ is an absolutely continuous function, $h_{j,k}\left( \tau \right)$ is given as:
\begin{align}\label{eq:h_discrete}
    \begin{split}
        &h_{j,k}(\tau)= \frac{\tau^{n+1-j}}{(n+1-j)} f_0^{(n+1)}(\theta_j), \\
        &\theta_j \in \left( t_k,t_{k+1} \right), \;\;\;\; j=0,1,2,\cdots, n.
    \end{split}
\end{align}

For a discontinuous function $f_0^{(n+1)}(t)$, $h_{j,k}(\tau)$ is presented as:
\begin{align}\label{eq:h_discrete_discont}
    \begin{split}
        &h_{j,k}(\tau)\in \frac{\tau^{n+1-j}}{(n+1-j)} \left[-1,1  \right], \\
        &j=0,1,2,\cdots, n.
    \end{split}
\end{align}

\subsection{Time Discretization of the Robust Exact Filtering Differentiator}\label{timeDiscretization}
For the differentiator \eqref{eq:cont_dist_filt_mat}, $z_j(t_{k+1})=z_{j,k+1}$ is proposed as a copy of $x_{j,k+1}$ with a injection term $\Gamma_{j+n_f+1,k}w_{1,k}$:

\begin{align}
    \begin{split}\label{eq:disc_z}
        z_{j,k+1}&=\sum_{l=j}^{n} \frac{\tau^{l-j}}{(l-j)!} z_{l,k}+\Gamma_{j+n_f+1,k}w_{1,k}.\\
        j&=0,1,2,\cdots,n.
    \end{split}
\end{align}

Evidently, $h_{j,k}(\tau)$ is omitted because it is not measured. Furthermore, $\tau$ is considered constant. $\Gamma_{j+n_f+1,k}$ is defined after. Based on Euler discretization, $w_{j,k+1}$ is proposed as:

\begin{align}
    \begin{split}\label{eq:disc_w}
        w_{j,k+1}  &= w_{j,k}  +\tau w_{j+1,k}+\Gamma_{j,k}w_{1,k},\\
        w_{n_f,k+1}&= w_{n_f,k}+\tau (z_{0,k}-f(t)) +\Gamma_{n_f,k}w_{1,k},\\
        i&=1,2,\cdots, n_f-1.
    \end{split}
\end{align}
where $\Gamma_{j,k}$ are defined after. Using the representations \eqref{eq:disc_z} and \eqref{eq:disc_w}, the discrete-time differentiator is summarized as:

\begin{align}
    \begin{split}\label{eq:disc_diff}
        \left[ \begin{array}{cc}
         \bm{w}_{k+1}  \\
         \bm{z}_{k+1} 
         \end{array} \right]=  \bm{\Psi}(\tau) \left[ \begin{array}{cc}
         \bm{w}_{k}  \\
         \bm{z}_{k} 
         \end{array} \right]-\tau \bm{e}_{n_f} f(t)
         +\bm{\Gamma}_k w_{1,k},
    \end{split}
\end{align}
where 
$\bm{w}_{k}=\left[w_{1,k} \;\;\; w_{2,k} \;\;\; \cdots \;\;\; w_{n_f,k}  \right]^T$, $\bm{z}_{k}=\left[z_{0,k} \;\;\; z_{1,k} \;\;\; \cdots \;\;\; z_{n,k}  \right]^T$, $\bm{\Gamma}_{k}=\left[\Gamma_{1,k} \;\;\; \Gamma_{1,k} \;\;\; \cdots \;\;\; \Gamma_{m+1,k}  \right]^T$, $\Psi(\tau)$ is given as:

\begin{align}
    \begin{split}
        \Psi(\tau)=\left[ \begin{array}{cccccccccc}
         1 & \tau & 0 &  \cdots & 0 & 0 & 0 & 0 & \cdots & 0\\
         0 & 1 & \tau &  \cdots & 0 & 0 & 0 & 0 & \cdots & 0\\
         \vdots & \vdots & \vdots & \cdots & \vdots & \vdots & \vdots & \vdots & \cdots & \vdots\\
         0 & 0 & 0 & \cdots & 1 & \tau & 0 & 0 & \cdots & 0\\
         0 & 0 & 0 & \cdots & 0 & 1 & \tau & \frac{\tau^2}{2!} & \cdots & \frac{\tau^n}{n!} \\
         0 & 0 & 0 & \cdots & 0 & 0 & 1 & \tau  & \cdots & \frac{\tau^{(n-1)}}{(n-1)!} \\
         \vdots & \vdots & \vdots & \cdots & \vdots & \vdots & \vdots & \vdots & \cdots & \vdots\\
         0 & 0 & 0 & \cdots & 0 & 0 & 0 & 0 & \cdots & 1\\
         \end{array} \right].
    \end{split}
\end{align}

Note that the first $n_f$ rows of $\bm{\Psi}(\tau)$ only present $1$, $0$ and $\tau$ and the dimension of $\bm{\Psi}(\tau)$ is $(m+1) \times (m+1)$. Equivalently to continuous-time system error, the discrete-time system error of \eqref{eq:disc_diff} can be represented as:

\begin{align}
    \begin{split}\label{eq:disc_diff_mat}
        \left[ \begin{array}{cc}
         \bm{w}_{k+1}  \\
         \bm{\sigma}_{k+1} 
         \end{array} \right]=  \left(\bm{\Psi}(\tau)+\bm{\Gamma}_k \bm{e}_{1}^T\right)\left[ \begin{array}{cc}
         \bm{w}_{k}  \\
         \bm{\sigma}_{k} 
         \end{array} \right]-\left[ \begin{array}{cc}
         \bm{0}  \\
         \bm{h}_{k}(\tau) 
         \end{array} \right]
        ,
    \end{split}
\end{align}
where $\bm{\sigma}=\left[ \sigma_{0,k}\;\; \sigma_{1,k} \;\; \cdots \;\; \sigma_{n,k}\right]$, $\bm{h}_k(\tau)=\left[ h_{0,k}(\tau)\;\; h_{1,k}(\tau) \;\; \cdots \;\; h_{n,k}(\tau)\right]$, $\bm{e}_1^T=\left[1 \;\; 0 \;\; \cdots \;\; 0 \right]$ and dimension of $\bm{e}_1$  is $(m+1) \times 1$. Let  $d_j$ the desired eigenvalues of the discrete-time system, then the desired characteristic polynomial is given as $\displaystyle P_d(r)= \prod_{j=1}^{m+1} \left(r-d_j \right)$ and for a matrix case $\displaystyle P_d(\bm{\Psi}(\tau))= \prod_{j=1}^{m+1} \left(\bm{\Psi}(\tau)-d_j \bm{I} \right)$. The desired characteristic polynomial evaluated in $\bm{\Psi}(\tau)+\bm{\Gamma}_k \bm{e}_1^T$ is given as $P_d(\bm{\Psi}(\tau)+\bm{\Gamma}_k \bm{e}_1^T)=(\bm{\Psi}(\tau)+\bm{\Gamma}_k\bm{e}_1^T)^{m+1}+\displaystyle  \sum_{j=0}^{m} \alpha_j (\bm{\Psi}(\tau)+\bm{\Gamma}_k \bm{e}_1^T)^{j}$. Here: 

\begin{align}
    \begin{split}
        (\bm{\Psi}(\tau)+\bm{\Gamma}_k\bm{e}_1^T)^{0}&=\bm{I}\\
        (\bm{\Psi}(\tau)+\bm{\Gamma}_k\bm{e}_1^T)^{1}&=(\bm{\Psi}(\tau)+\bm{\Gamma}_k\bm{e}_1^T)\\
        (\bm{\Psi}(\tau)+\bm{\Gamma}_k\bm{e}_1^T)^{2}&=\bm{\Psi}^2(\tau)+\bm{\Gamma}_k\bm{e}_1^T\bm{\Psi}(\tau)+ \ldots\\
        &\;\;\;\;\; \ldots +\bm{\Psi}(\tau)\bm{\Gamma}_k\ e_1^T+ \bm{\Gamma}_k\bm{e}_1^T\bm{\Gamma}_k\bm{e}_1^T\\
                                               & \;\; \vdots\\
        (\bm{\Psi}(\tau)+\bm{\Gamma}_k\bm{e}_1^T)^{m+1}&=\bm{\Psi}(\tau)^{m+1} + \bm{\Gamma}_k \bm{e}_1^T \bm{\Psi}(\tau)^{m}+\cdots.\\
    \end{split}
\end{align}

Therefore we obtain the following equation:

\begin{align}
    \begin{split}
        P_d(\bm{\Psi}(\tau)+\bm{\Gamma}_k \bm{e}_1^T)=P_d(\bm{\Psi}(\tau))+\left[ * \;\;\; \cdots \;\;\ * \;\; \bm{\Gamma}_k \right] \bm{S}.
    \end{split}
\end{align}

Due to the Cayley–Hamilton theorem $ P_d(\bm{\Psi}(\tau)+\bm{\Gamma}_k \bm{e}_1^T)=\bm{0}$ and, therefore, $\bm{\Gamma}_k$ can be calculated as:

\begin{align}\label{eq:gamma}
    \begin{split}
        \bm{\Gamma}_k=- P_d(\bm{\Psi}(\tau)) \bm{S}^{-1}\bm{e}_{m+1},
    \end{split}
\end{align}
where

\begin{align}
    \begin{split}
       \bm{S}= \left[ \begin{array}{cc}
         \bm{e}_{1}^T  \\
         \bm{e}_{1}^T \Psi(\tau) \\ 
         \bm{e}_{1}^T \Psi^2(\tau) \\ 
         \vdots \\
         \bm{e}_{1}^T \Psi^m(\tau) \\ 
         \end{array} \right] 
    \end{split}
\end{align}

Now, the objective is to select adequate roots $d_j$. With the purpose of emulate the behavior of the continuous-time system, a mapping of the continuous-time domain to the discrete-time domain is used, Euler $\left( d_j=1+\tau c_j\right)$, Matching $\left( d_j=e^{\tau c_j} \right)$ and Bilinear $\left( d_j=\frac{1+c_j \tau/2 }{1-c_j \tau/2}\right)$ approach to name a few. As $c_j=|w_1|^{\frac{-1}{m+1}} b_j$, Euler and Bilinear approaches have a singularity at $w_{1}=0$, consequently, a Matching approach is used:

\begin{align}\label{eq:cont_roots_mat}
\begin{split}
    d_j=e^{\tau c_j}=e^{\tau |w_1|^{\frac{-1}{m+1}} b_j}.
\end{split}
\end{align}

\begin{thm}\label{th:1}
Let the discrete-time differentiator \eqref{eq:disc_diff} with $\bm{\Gamma}_k$ defined as \eqref{eq:gamma}, $d_j$ defined as \eqref{eq:cont_roots_mat}, $|f_0^{n+1}(t)|\leq L$ and $\Delta(t)=0$. If $Real\left(b_j\right)<0$ then the trajectories of the error system \eqref{eq:disc_diff_mat} converge to a neighborhood of the origin and keep in this neighborhood, which is defined as:

\begin{align}
    \begin{split}
       \norm{ \left[ \begin{array}{cc}
         \bm{w}_{k}  \\
         \bm{\sigma}_{k} 
         \end{array} \right] }_2 \leq  K  \norm{ \bm{h}_k\left( \tau \right) }_2.
    \end{split}
\end{align}
\end{thm}

$K$ and the proof are presented in the Appendix \ref{pr:theorem1}. Note that the roots $b_j$ can be selected independent of $\lambda_j$ and $L$ because of Theorem \ref{th:1}. This allows to implement the differentiator even if $L$ is unknown. Furthermore, if $b_j$ are selected as $b_1=b_2=b_3=\cdots=b_{m+1}$, $\bm{\Gamma}_k$ presents a less complex equation than for $b_j \neq b_{j+1}$.

\section{Results}\label{sec:Results}

In this Section, two simulations are performed. In the first one, a free-noise case is considered. To implement \eqref{eq:disc_diff_mat}, $\bm{\Gamma}_k$ is calculated offline and expressed  as a function of $d_j$, $d_j$ is updated using Equation \eqref{eq:cont_roots_mat}. Four differentiators are considered, three of them with repeated $b_j$ and one where its roots corresponds to the roots of the polynomial \eqref{eq:pol_conti_b}, where $\lambda_j$ are selected as in \cite{levant2019robust} and $L$ is selected as $|f_0(t)|\leq L$. For the last one, $b_j$ will be represented as $b_j(L,\lambda)$.

\subsection{Simulation I}

\begin{figure}[H]
	\includegraphics[width=0.51\textwidth]{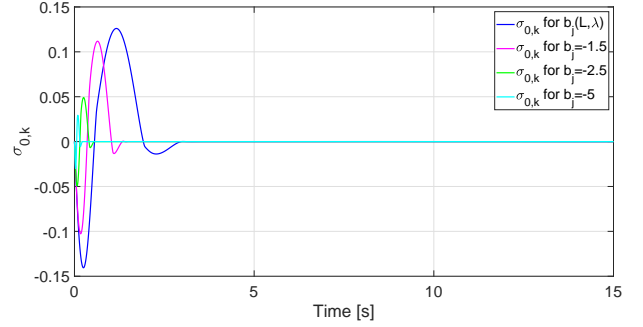} 
	\caption{Estimation Error of $f_0(t)$.}
	\label{Fig:Function}
\end{figure}

\begin{figure}[H]
	\includegraphics[width=0.51\textwidth]{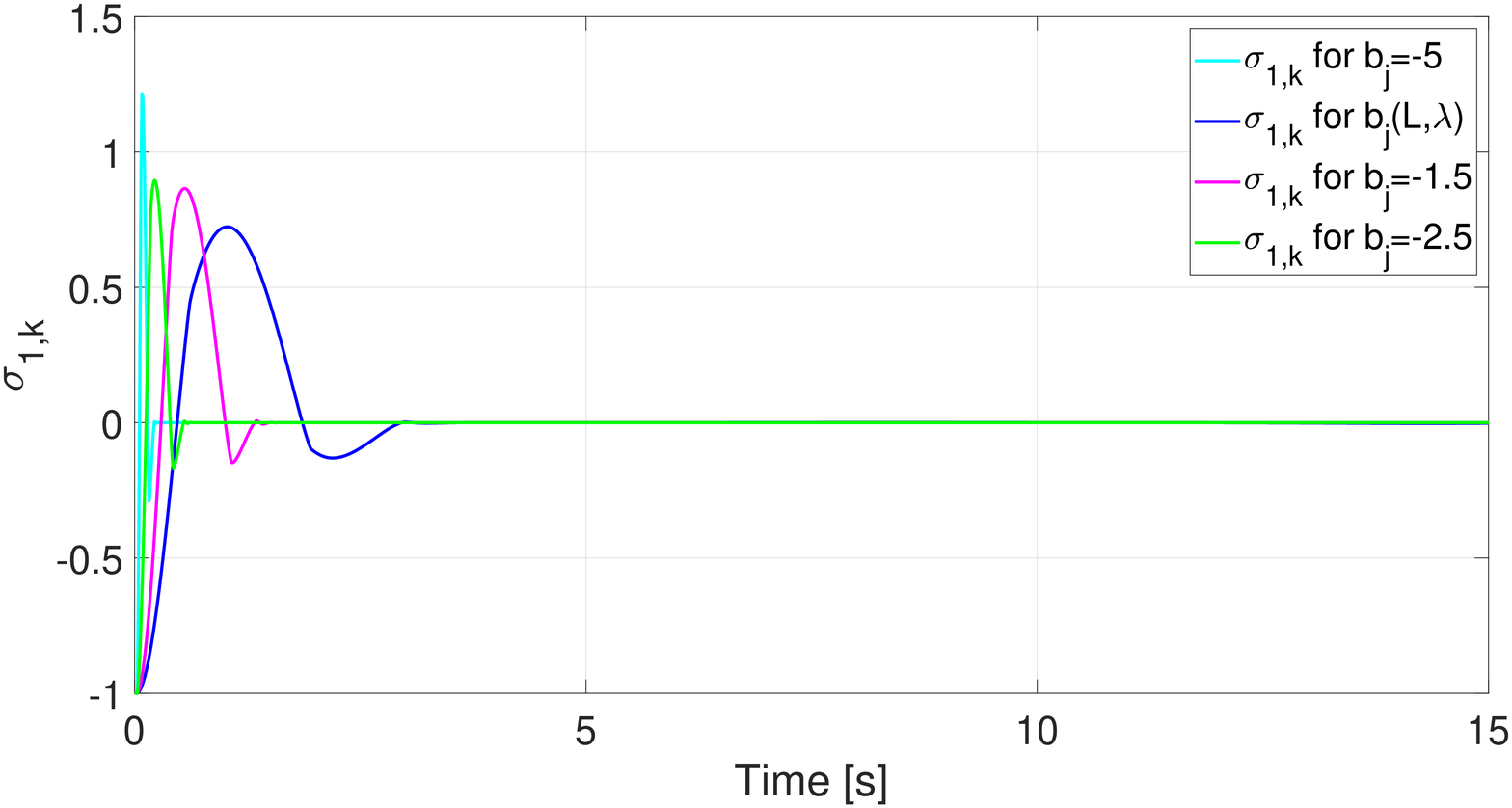}
	\caption{Estimation Error of $f_0^{(1)}(t)$.}
	\label{Fig:Function1}
\end{figure}

\begin{figure}[H]
	\includegraphics[width=0.51\textwidth]{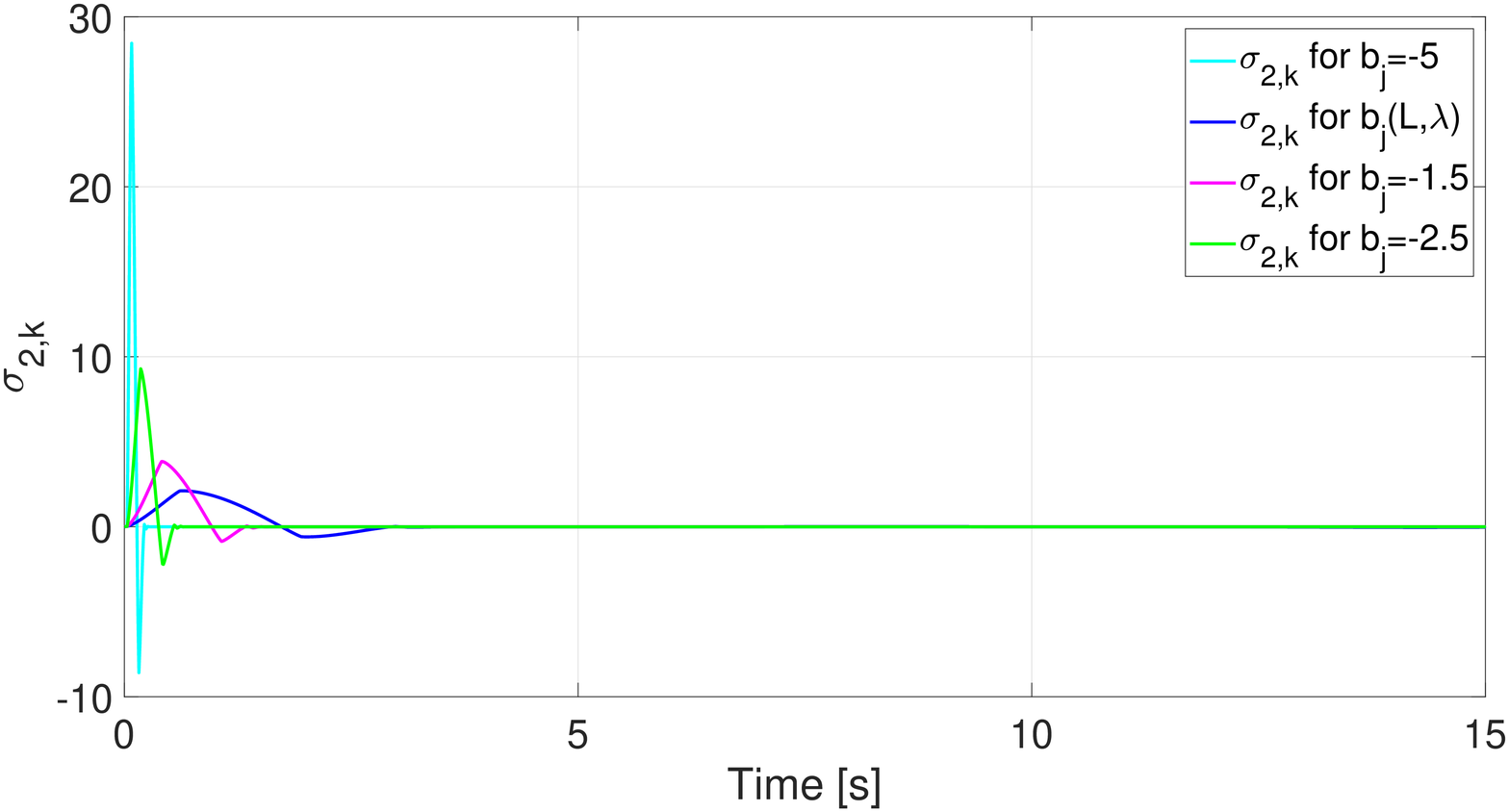}
	\caption{Estimation Error of $f_0^{(2)}(t)$.}
	\label{Fig:Function2}
\end{figure}

\begin{figure}[H]
	\includegraphics[width=0.51\textwidth]{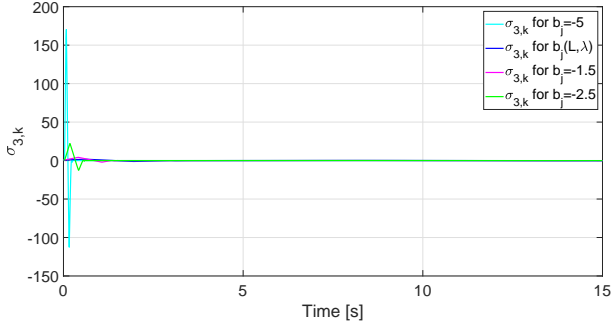}
	\caption{Estimation Error of $f_0^{(3)}(t)$.}
	\label{Fig:Function3}
\end{figure}

Here, a free-noise case is considered, $n_f=2$, $n=3$, $\tau=0.01 \sec$, $\lambda_0=1.1$, $\lambda_1=6.75$,  $\lambda_2=20.26$, $\lambda_3=32.24$, $\lambda_4=23.72$ and $\lambda_5=7$. For this simulation, $f_0(t)=t \cos{(t/2)}$, then $|f^{(4)}(t)|\leq L=2$ for $t \leq 31.54619 \sec$, $b_1(L,\lambda)=-2.8072 + 2.7583i$, $b_2(L,\lambda)=-2.8072 - 2.7583i$, $b_3(L,\lambda)=-0.2725 + 0.3729i$, $b_4(L,\lambda)=-0.2725 - 0.3729i$ , $b_5(L,\lambda)=-1.0831$ and $b_6(L,\lambda)=-0.6148$. For the differentiator with repeated $b_j$, the selected roots are $b_j=-1.5$, $b_j=-2.5$, and $b_j=-5$. The estimation errors are presented in Figures \ref{Fig:Function}-\ref{Fig:Function3}. It can be seen that the differentiator presents an adequate estimation of the function and its derivatives. An interesting result is that the trajectories of the differentiators converge to a neighborhood with a different settling-time, this fact shows that there is a relation between this settling-time and the roots $b_j$.

\subsection{Simulation II}
\begin{figure}[H]
	\includegraphics[width=0.51\textwidth]{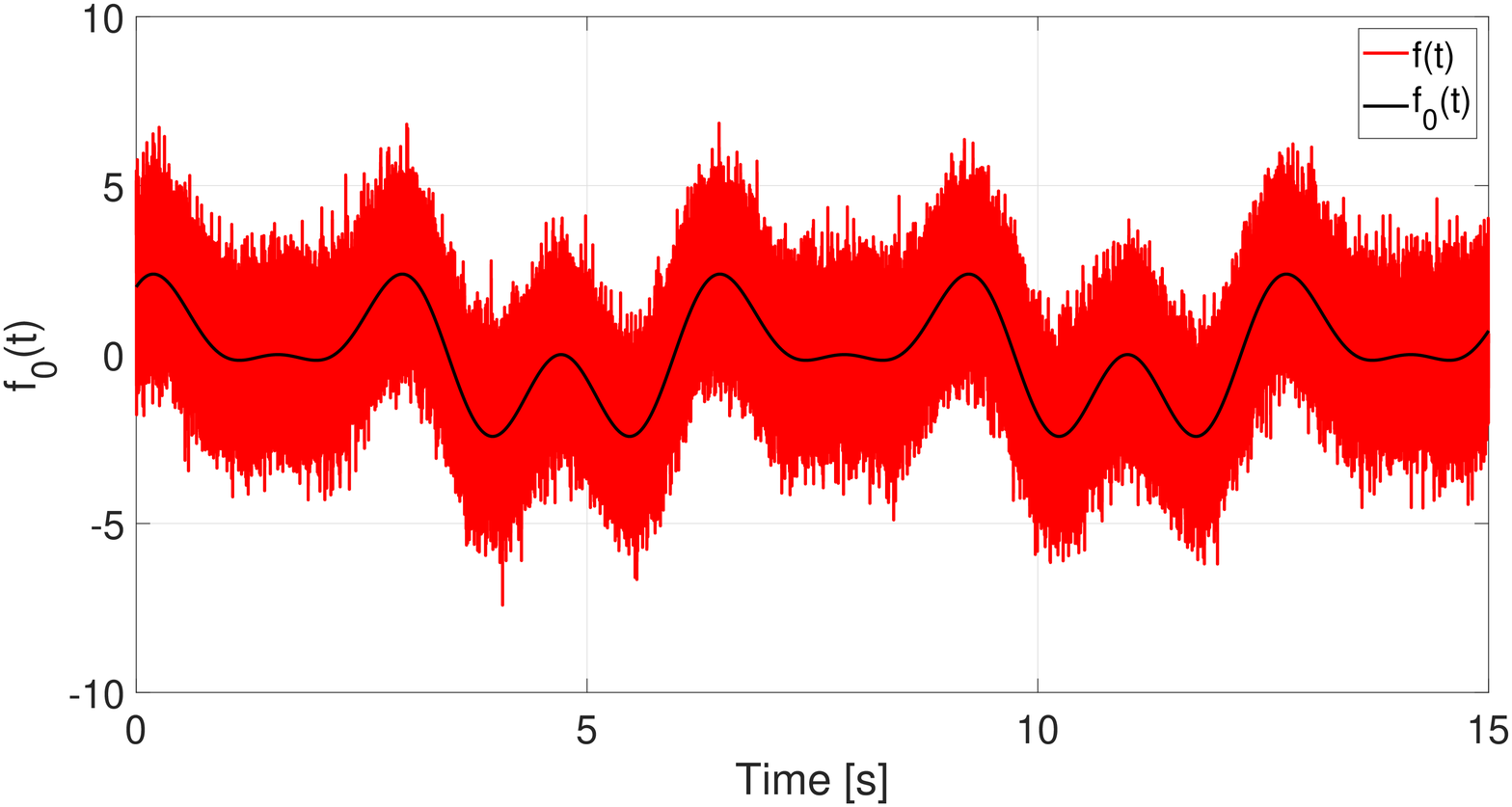}
	\caption{Estimation of $f_0(t)$.}
	\label{Fig:InputN}
\end{figure}

\begin{figure}[H]
	\includegraphics[width=0.51\textwidth]{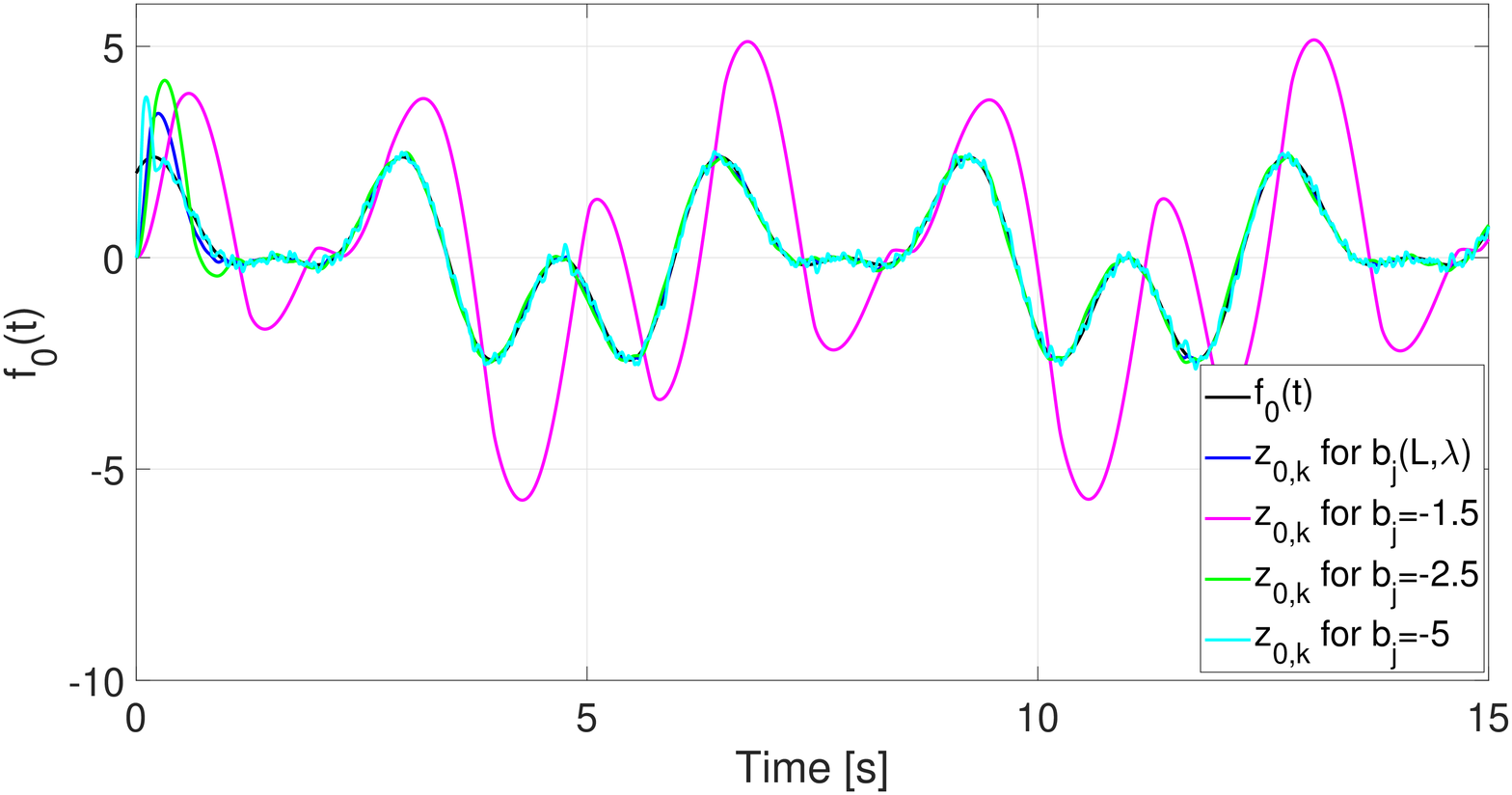}
	\caption{Estimation of $f_0(t)$.}
	\label{Fig:FunctionN}
\end{figure}

\begin{figure}[H]
	\includegraphics[width=0.51\textwidth]{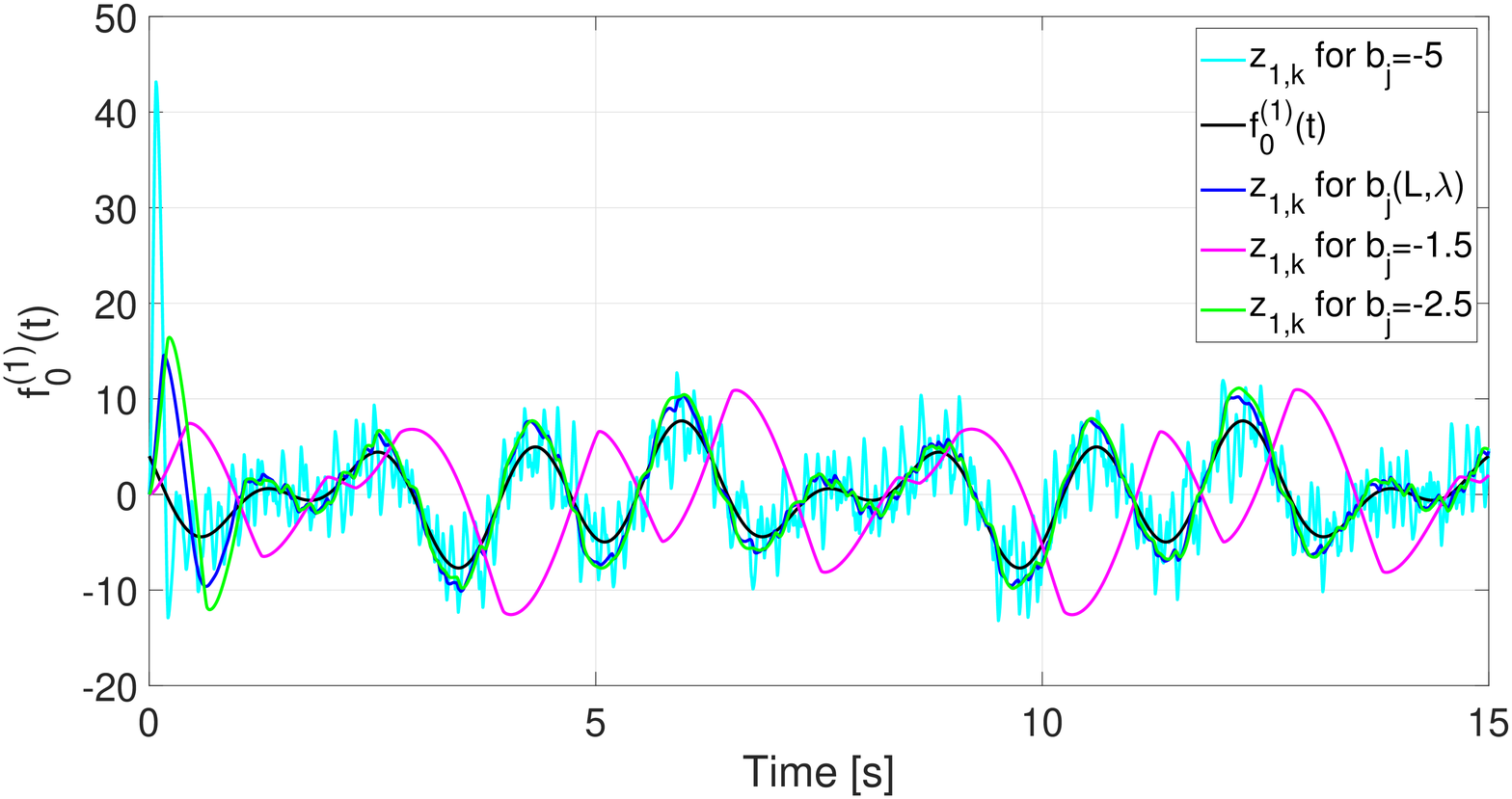}
	\caption{Estimation of $f_0^{(1)}(t)$.}
	\label{Fig:FunctionN1}
\end{figure}

\begin{figure}[H]
	\includegraphics[width=0.51\textwidth]{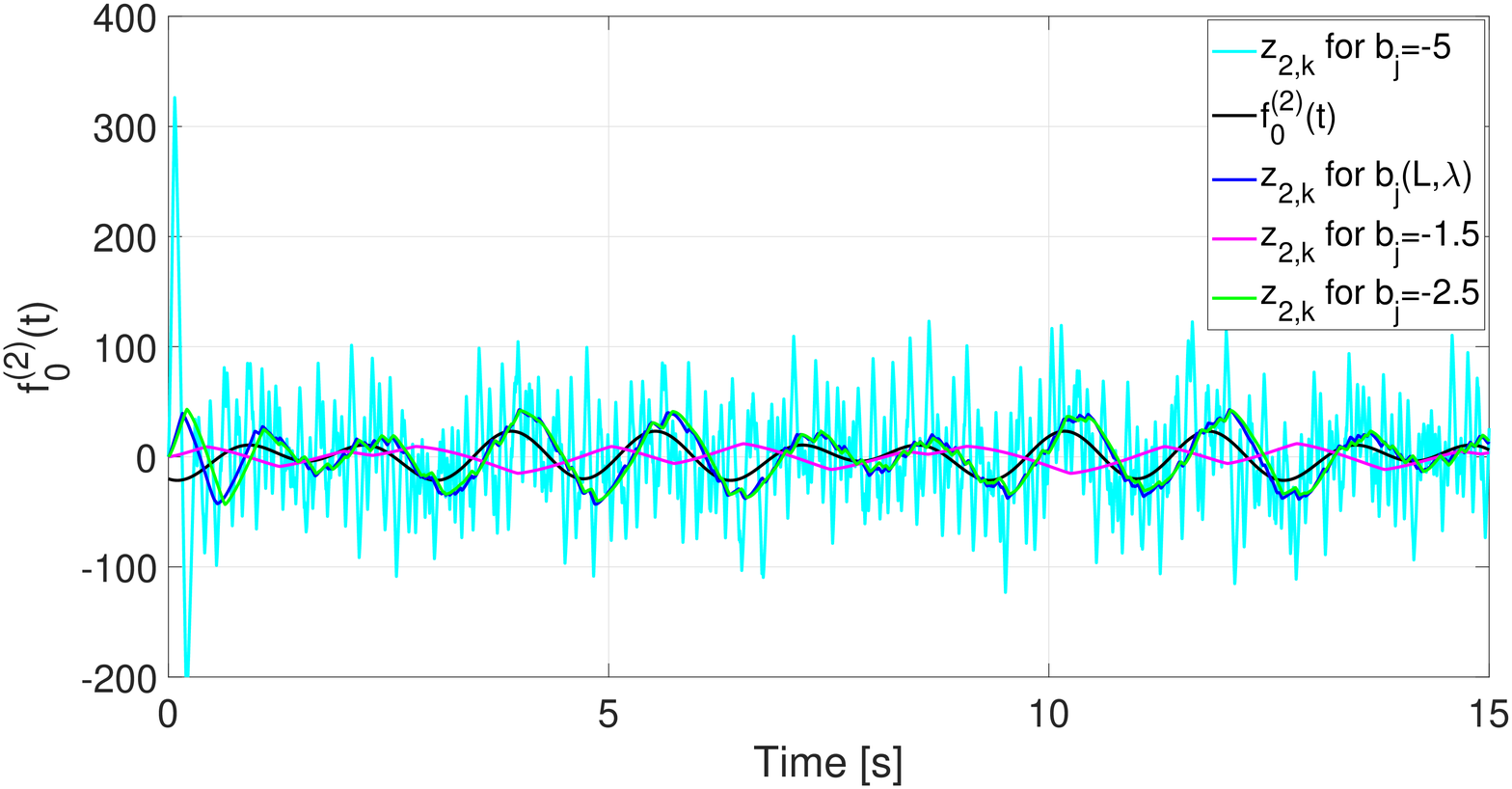}
	\caption{Estimation of $f_0^{(2)}(t)$.}
	\label{Fig:FunctionN2}
\end{figure}

\begin{figure}[H]
	\includegraphics[width=0.51\textwidth]{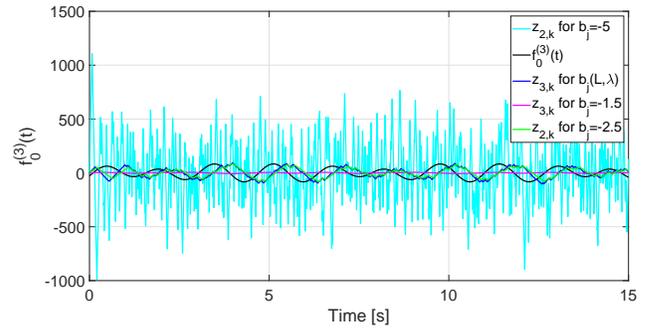}
	\caption{Estimation of $f_0^{(3)}(t)$.}
	\label{Fig:FunctionN3}
\end{figure}

In contrast to Simulation I, $\Delta(t)=\cos (10000 t)+\eta(t)$, $\eta(t)\sim\text{i.i.d. }\mathcal{N}(0,1^2)$, $f_0(t)=\sin (t)+\cos (2t) + \sin (3t) + \cos (4t)$, $|f^{(4)}(t)|\leq L=320$, $b_1(L,\lambda)=-6.5408 + 6.4269i$, $b_2(L,\lambda)=-6.5408 - 6.4269i$, $b_3(L,\lambda)=-0.6348 + 0.8689i$, $b_4(L,\lambda)=-0.6348 - 0.8689i$ , $b_5(L,\lambda)=-2.5235$ and $b_6(L,\lambda)=-0.6348$ are considered. The repeated poles are as in Simulation I.  Furthermore, a lower sampling time is used to obtain adequate estimations, $\tau=0.0001 \sec$. The noisy input and $f_0(t)$ are presented in Figure \ref{Fig:InputN} whereas the estimation $z_{j,k}$ are shown in Figures \ref{Fig:FunctionN}-\ref{Fig:FunctionN3}. 

\begin{figure}[H]
	\includegraphics[width=0.51\textwidth]{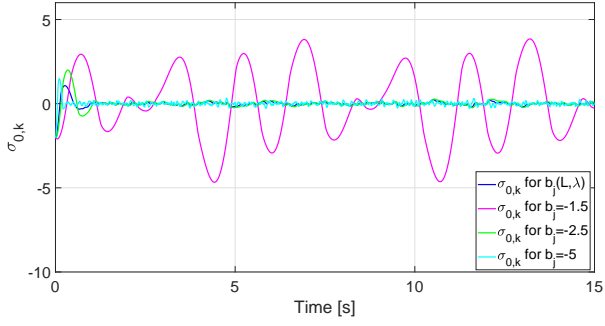}
	\caption{Estimation Error of $f_0(t)$.}
	\label{Fig:EFunctionN}
\end{figure}

\begin{figure}[H]
	\includegraphics[width=0.51\textwidth]{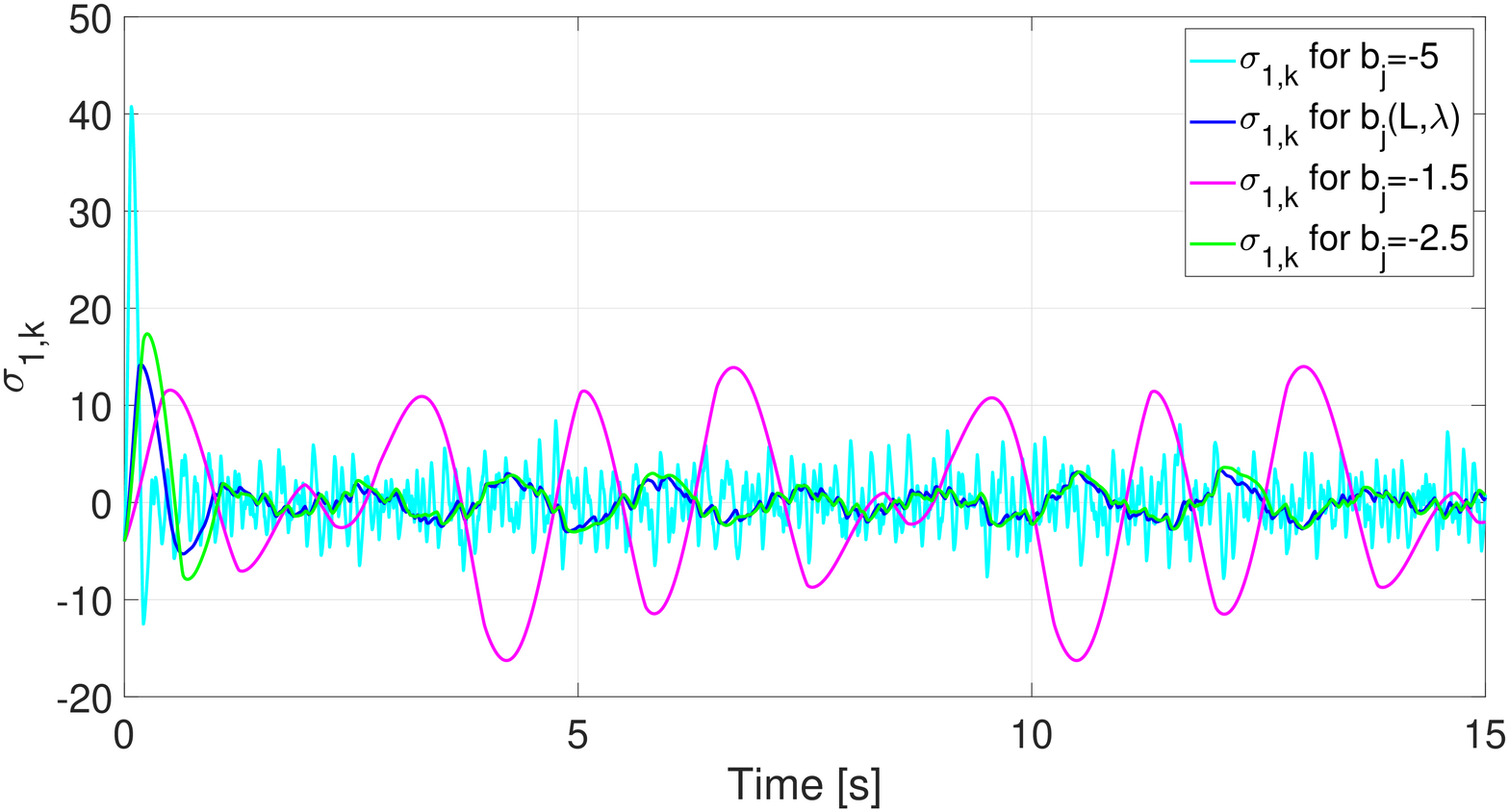}
	\caption{Estimation Error of $f_0^{(1)}(t)$.}
	\label{Fig:EFunctionN1}
\end{figure}

\begin{figure}[H]
	\includegraphics[width=0.51\textwidth]{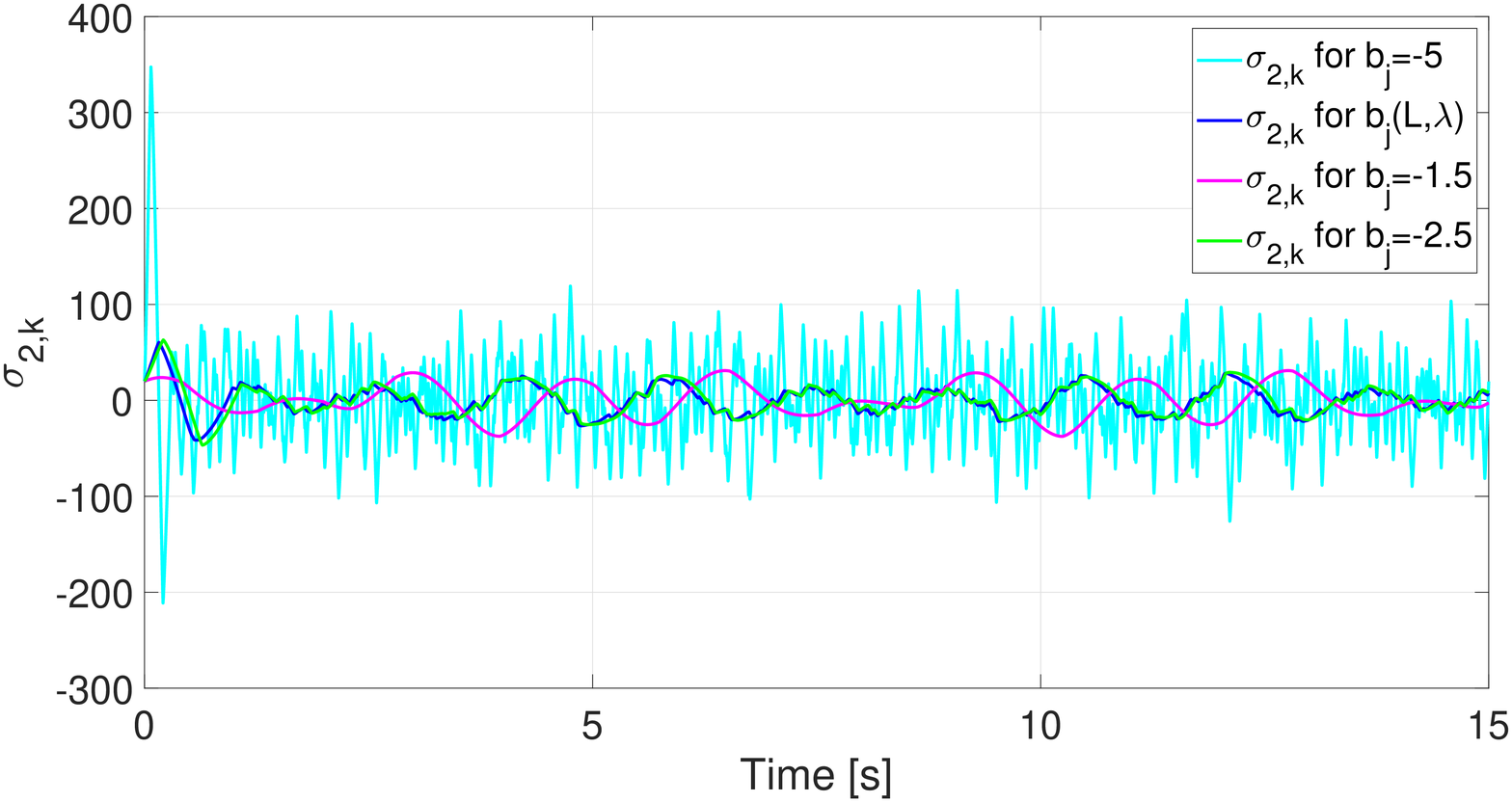}
	\caption{Estimation Error of $f_0^{(2)}(t)$.}
	\label{Fig:EFunctionN2}
\end{figure}

\begin{figure}[H]
	\includegraphics[width=0.51\textwidth]{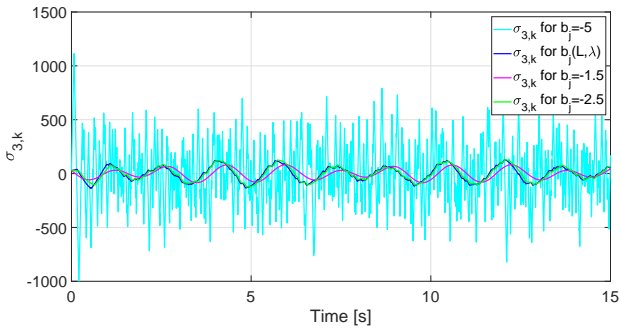}
	\caption{Estimation Error of $f_0^{(3)}(t)$.}
	\label{Fig:EFunctionN3}
\end{figure}

Note that $\cos (10000 t)$ can be represented as a signal of global filtering $j$ for any integer $j\geq 0$, the accuracy \eqref{eq:acc_robust_dif} is better than \eqref{eq:acc_standard_dif}. As it can be seen in Figure \ref{Fig:EFunctionN}-\ref{Fig:EFunctionN3}, the best estimations come from the differentiators with $b_j=-2.5$ and $b_j=(L,\lambda)$. Although the noisy signal presented in Figure \ref{Fig:FunctionN}, both differentiators present adequate estimations of the function $f_0(t)$ and its derivatives. It is interesting that for repeated real $b_j$, a low $b_j$ increase the sensitivity to noise whereas a high $b_j$ reduces its accuracy.

\section{Conclusion}\label{sec:conclusion}

A new time discretization of the robust exact filtering differentiator \eqref{eq:cont_dist_filt} is presented. It can be implemented with or without the knowledge of $L$ and $\lambda_j$. It was demonstrated that, for a free-noise case, and, under the assumption of Theorem \ref{th:1}, the trajectories of the system converge to a neighborhood of the origin. The Simulation I suggests a relation between the settling-time to the neighborhood of Theorem \ref{th:1} and the roots $b_j$, whereas Simulation II shows a relation between sensitivity to noise and $b_j$. Future work will address a demonstration of the convergence of the error system's trajectories to a neighborhood of the origin in the presence of a noisy input and estimation of this neighborhood and its respective settling-time function.


\appendix

\section{Appendix. Proof of Theorem 1} \label{pr:theorem1}

Let $\bm{E}=\left(\bm{\Psi}(\tau)+\bm{\Gamma}_k \bm{e}_1^T  \right)$ and the Lyapunov function: 

\begin{align}\label{eq:Lyapunov}
    \begin{split}
        V_k=\left[ \begin{array}{cc}
         \bm{w}_{k}  \\
         \bm{\sigma}_{k} 
         \end{array} \right]^T \bm{P} \left[ \begin{array}{cc}
         \bm{w}_{k}  \\
         \bm{\sigma}_{k} 
         \end{array} \right],
    \end{split}
\end{align}
where $\bm{P}$ is a real symmetric positive definite matrix and it is such that 

\begin{align}
    \begin{split}
        \bm{E}^T \bm{P} \bm{E} -\bm{P}=-\bm{Q},
    \end{split}
\end{align}
$\bm{Q}$ is a real symmetric positive definite matrix and $\lambda_{\min}(Q)>1$. From Equations \eqref{eq:disc_diff_mat} and \eqref{eq:Lyapunov}:  

\begin{align}
    \begin{split}
         &V_{k+1}-V_k=\\
         &= -\left[ \begin{array}{cc}
         \bm{w}_{k}  \\
         \bm{\sigma}_{k} 
         \end{array} \right]^T \bm{Q} \left[ \begin{array}{cc}
         \bm{w}_{k}  \\
         \bm{\sigma}_{k} 
         \end{array} \right] +\left[ \begin{array}{cc}
         \bm{0} \\
         \bm{h}_{k}(\tau) 
         \end{array} \right]^T \bm{P} \left[ \begin{array}{cc}
         \bm{0} \\
         \bm{h}_{k}(\tau) 
         \end{array} \right] \\
         & \;\;\; \;-  2 \left[ \begin{array}{cc}
         \bm{w}_{k}  \\
         \bm{\sigma}_{k} 
         \end{array} \right]^T \bm{E}^T \bm{P} \bm{E} \left[ \begin{array}{cc}
         \bm{0} \\
         \bm{h}_{k}(\tau) 
         \end{array} \right] .
    \end{split}
\end{align}
By using inequality \eqref{eq:matrixprop}, the following inequality is obtained:

\begin{align}
    \begin{split}
         &V_{k+1}-V_k\leq  (\lambda_{\max}(\bm{E})+\lambda_{\max}\left( \bm{P}\right))  \norm{\bm{h}_k(\tau)}_2^2-\cdots\\
         &-\left(\lambda_{min}(Q)-1\right) \norm{ \left[ \begin{array}{cc}
         \bm{w}_{k}  \\
         \bm{\sigma}_{k} 
         \end{array} \right]}_2^2.
    \end{split}
\end{align}
Therefore with the condition
\begin{align}
    \begin{split}
       &\norm{ \left[ \begin{array}{cc}
         \bm{w}_{k}  \\
         \bm{\sigma}_{k} 
         \end{array} \right] }_2 >  K  \norm{ \bm{h}_k\left( \tau \right) }_2,\\
        & K=\sqrt{  \frac{\lambda_{\max}(\bm{E})+\lambda_{\max}(\bm{P})}{\lambda_{\min}(\bm{Q}) -1 }  },
    \end{split}
\end{align}
one obtains that $V_{k+1}-V_k<0$. \QEDA

\bibliographystyle{IEEEtran}
\bibliography{IEEEexample}

\end{document}